\documentstyle{l-aa}

\def\Lya{\mbox{Ly$\alpha$}} 
\def\Ha{\mbox{H$\alpha$}}
\def\Hb{\mbox{H$\beta$}}
\def\CIV{\mbox{C${\sc IV}\lambda$1550}}
\def\HeII{\mbox{He${\sc II}\lambda$1640}}
\def\NV{\mbox{N${\sc V}\lambda$1240}}

\def\Oiii{\mbox{O${\sc III}\lambda\lambda$4959,5007}}
\def\Nii{\mbox{N${\sc II}\lambda\lambda$6548,6584}}
\def\chicua{\mbox{$\chi^{2}$}}
\def\chicuar{\mbox{$\chi^{2}_{\nu}$}}

\def\ergcms{\,erg\,cm$^{-2}$\,s$^{-1}$}
\def\ergs{\,erg\,s$^{-1}$}
\def\yo{Pedro M. Rodr\'{\i}guez--Pascual}
\def\kms{\mbox{km s$^{-1}$}}
\def\afir{\mbox{$\alpha(25,60)$}}

\def\iue{{\it IUE}}
\def\rosat{{\it ROSAT}}
\def\iras{{\it IRAS}}
\def\Luv{\mbox{$L_{UV}$}}
\def\Lx{\mbox{$L_{ROSAT}$}}
\def\L60{\mbox{$L_{60\mu m}$}}
\def\irastrece{\mbox{IRAS13224-3809}}

\input epsf
\begin{document}

\thesaurus{03(11.14.1; 11.19.1; 13.21.1; 13.25.2)}

\title{The Broad Line Region of Narrow--line Seyfert 1 galaxies}
\subtitle{}

\author{\yo\inst{1}\thanks{Affiliated to the Astrophysics Division, SSD,
ESTEC} \and J. Miguel Mas--Hesse\inst{2} \and Mar\'{\i}a 
Santos--Lle\'o\inst{2}} 
\offprints{\yo} 
\institute{ESA - IUE Observatory, P.O. Box 50727, 28080 Madrid, Spain \and
LAEFF--INTA, P.O. Box 50727, 28080 Madrid, Spain}

\date{Received March 19; accepted June 9, 1997}

\maketitle

\begin{abstract}

We have analyzed new and archival {\it IUE} observations of narrow-line
Seyfert~1 galaxies (NLS1) in order to revise the ultraviolet (UV)
properties of this sub-group of Active Galactic Nuclei (AGN). We have found
broad wings in the strongest UV emission lines, ruling out the hypothesis
that there is no broad line emission region in this type of objects. Since
the similarities in spectral energy distributions from the far-infrared
(FIR) to the soft X rays in both narrow-line and broad-line Seyfert~1
galaxies do not suggest that the nuclei of NLS1 are hidden from a direct
view, we discuss the possibility that the line emitting material in NLS1 is
optically thin.


\end{abstract}

\section{Introduction}

The class of narrow--line Seyfert 1 (NLS1) galaxies is characterized by the
properties of their optical emission lines (Osterbrock \& Pogge 1985;
Goodrich 1989): the permitted lines are narrower ($\sim $1\,000 \kms full
width at half-maximum, FWHM) than in typical Seyfert~1 galaxies ($\sim $5\,000
\kms, FWHM), though still slightly broader than the forbidden lines; the
[OIII]/\Hb\ ratio is much smaller than in Seyfert~2; narrow lines of highly
ionized atoms are detected; the optical FeII emission is among the
strongest detected in active galactic nuclei (AGN).

The radio properties of NLS1 galaxies (Ulvestad, Antonucci \&\ Goodrich
1995) are similar to those found in other Seyfert galaxies. The only
exception refers to the orientation of the radio axes with respect to the
optical polarization: Mrk~766 and Mrk~1126 are the only two known Seyfert
galaxies where the radio axis is perpendicular to the
polarization. Spectropolarimetric observations of a sample of 17 NLS1 galaxies
(Goodrich 1989) revealed that 6 of them show significant intrinsic
polarization.

It has been claimed (Boller et al. 1996 and references therein) that in the
{\it ROSAT} band, NLS1 show both steeper spectra and stronger variability 
than typical broad line
Seyfert 1 galaxies. Moreover, the soft X-ray 
variability time scales found in some NLS1 are among the fastest of Seyfert
galaxies: the doubling times are shorter than 15000 seconds for 4 out of 30
NLS1. In spite of the different spectral shapes, the soft X--ray
luminosities of NLS1 are similar to those of typical broad line
Seyfert~1s. At higher energies, only three objects have been reported to be
observed so far: IRAS~13224-3809 shows a hard ($\Gamma\sim$\,1.3) power law
from 2 to 10 keV while below 2 keV the spectrum is dominated by a soft
excess (Otani 1995); in contrast, RE\,1034+39 shows a very steep spectrum
($\Gamma\sim$\,2.6) in the range 2--10 keV (Pounds, Done \&\ Osborne 1995);
finally, Leighly et al. (1996) find rapid variations (doubling time scale
$\sim$1\,000 seconds) in the {\it ASCA} data (0.4--10.5keV) of
Mrk~766. These authors model the combined {\it ROSAT}-{\it ASCA} spectra of
Mrk~766 with a power law, warm absorber and soft excess. With this model,
the spectral variability observed can be attributed to changes in the
spectral index of the power law and in the ionization degree 
of the warm absorber.

The narrowness of the optical permitted lines has been interpreted as an
orientation effect of a disc shaped BLR (Osterbrock \& Pogge 1985; Goodrich
1989; Stephens 1989; Puchnariewitz et al. 1992). If the clouds are confined
to move in a plane, the emission lines will be much narrower when viewed
from a line of sight nearly perpendicular to the plane. In this pole--on
model, the rapid X-ray variability might be understood in terms of relativistic
beaming effects if we were looking down an outflow from the central
engine. If such beaming effects are not present, a rapid intrinsic
variability can only result from a very compact emitting region. 

Another possibility, that does not consider orientation effects as playing a
major role, is based on the steep spectrum generally observed in the soft
X--rays. If it is the high energy tail of the emission from an accretion
disc, such ``hot'' spectrum will be produced when the central black hole
has a relatively small mass compared to those in normal Seyfert~1's (Ross
\& Fabian 1993). In this scenario, the rapid variability observed in the
X--rays would be naturally explained by the smaller dimension of the
emitting region. Moreover, if the gravitational force from the central
black hole dominates the BLR kinematics, smaller black holes would
result in lower cloud velocities, assuming these clouds are kept at
distances similar to those in normal Seyfert~1's. This situation may occur
if the larger ionization parameter (the ratio of the density of ionizing
photons to hydrogen atoms), implied by the harder accretion disc
spectrum, hinders the formation of BLR clouds close to the central source.

It is remarkable that all these hypotheses about the nature of NLS1 are
based on the narrowness of the optical permitted lines. However, it is
known that the emission lines that better trace the innermost regions of
the BLR lie in the UV domain (Peterson 1994, and references
therein). Crenshaw et al. (1991) have reported {\it IUE} observations of
three NLS1 and found that the line ratios are similar to those found in
normal Seyfert~1 galaxies, but nothing is mentioned about the line widths.

In the very last years, {\it IUE} has observed a number of NLS1 galaxies,
increasing noticeably the sample discussed by Crenshaw et
al. (1991). Moreover, most of the results on the X-ray properties of NLS1
have come out also in the current decade. Our aim is then to revise the UV
properties of NLS1, based on a larger sample than in previous works, and
link these properties to those in other spectral ranges in order to
constrain the models currently proposed for this sub-class of AGN.

In Sect.~2 we describe the analyzed sample as well as its UV properties. In
this section we also combine the continuum measurements with far-infrared
(FIR) fluxes and soft X-ray properties from the literature to study the
spectral energy distribution of NLS1. The implications of the emission
lines analysis on the emitting-gas conditions are discussed in Sect~3 and
the main results are finally summarized in Sect.~4.

\begin{table*}

\caption[ ]{Log of IUE observations}
\label{log}
\begin{flushleft}

\begin{tabular}{lccc}
\multicolumn{1}{c}{Object} & 
\multicolumn{1}{c}{Image} & 
\multicolumn{1}{c}{Obs. date} & 
\multicolumn{1}{c}{Exp. Time} \\
 & & & \multicolumn{1}{c}{Min.} \\
\hline
             MRK 1044 & SWP56260 &   01/12/95 &  568 \\
             MRK1044  & SWP56319 &   20/12/95 &  225 \\
         KUG 1031+398 & SWP52918 &   27/11/94 &  280 \\
         KUG 1031+398 & SWP53041 &   11/12/94 &  310 \\
       IRAS13224-3809 & SWP46830 &   27/01/93 &  345 \\
       IRAS13224-3809 & SWP46914 &   11/02/93 &  340 \\
       IRAS13224-3809 & SWP47720 &   24/05/93 &  395 \\
       IRAS13224-3809 & SWP51149 &   21/06/94 &  400 \\
       IRAS13224-3809 & SWP54161 &   17/03/95 &  384 \\
       IRAS13224-3809 & SWP54216 &   24/03/95 &  369 \\
       IRAS13224-3809 & SWP55026 &   17/06/95 &  380 \\
       IRAS13224-3809 & SWP56586 &   17/01/96 &  360 \\
       IRAS13224-3809 & SWP56631 &   23/01/96 &  357 \\
       IRAS13224-3809 & SWP56646 &   29/01/96 &  330 \\
       IRAS13224-3809 & SWP56792 &   11/02/96 &  330 \\
\hline
\end{tabular}

\end{flushleft}
\end{table*}

\begin{table*}

\caption[ ]{Continuum Properties}
\label{cont}

\begin{flushleft}

\begin{tabular}{lcrrrrrcrrrr}
\hline
\multicolumn{1}{c}{Object} & 
\multicolumn{1}{c}{Redshift} & 
\multicolumn{1}{c}{$\nu F_{\nu}$(100$\mu$m)} &
\multicolumn{1}{c}{$\nu F_{\nu}$(60$\mu$m)} &
\multicolumn{1}{c}{$\nu F_{\nu}$(25$\mu$m)} &
\multicolumn{1}{c}{$\nu F_{\nu}$(12$\mu$m)} &
\multicolumn{1}{c}{$\nu F_{\nu}$(1450\AA)} &
\multicolumn{1}{c}{F(0.1-2.4keV)} &
\multicolumn{1}{c}{$\Gamma_{ROSAT}$} \\
 & & \multicolumn{6}{c}{10$^{-11}$\,erg\,cm$^{-2}$\,s$^{-1}$} \\
\hline
      MRK957 &  0.075 &  9.62$\pm$  0.96 &  10.47$\pm$  0.84 &   2.94$\pm$  0.38 &  $<$4.70 &   0.42$\pm$  0.10 &   0.36 &  3.0$\pm$0.2 \\
        IZW1 &  0.060 &  7.90$\pm$  0.79 &  11.22$\pm$  0.90 &  14.53$\pm$  1.45 &  12.80$\pm$  1.15 &   3.21$\pm$  0.19 &   2.51 &  3.0$\pm$0.1 \\
      MRK359 &  0.017 &  5.22$\pm$  0.68 &   5.66$\pm$  0.40 &   5.25$\pm$  1.16 &   2.98$\pm$  0.89 &   2.54$\pm$  0.12 &   2.95 &  2.4$\pm$0.1 \\
     MRK1044 &  0.016 &  2.64$\pm$  0.45 &   2.15$\pm$  0.22 &   2.58$\pm$  0.39 &   2.51$\pm$  0.63 &   3.84$\pm$  0.10 &   5.67 &  3.0$\pm$0.1 \\
     MRK1239 &  0.019 & $<$7.24 &   6.68$\pm$  0.53 &  13.69$\pm$  0.96 &  16.25$\pm$  1.79 &   0.24$\pm$  0.16 &   0.27 &  3.9$\pm$0.3 \\
 KUG1031+398 &  0.042 &  1.97$\pm$  0.41 &   1.74$\pm$  0.21 &  $<$2.10 &   3.35$\pm$  0.80 &   0.37$\pm$  0.10 &   5.16 &  4.4$\pm$0.1 \\
       MRK42 &  0.024 & $<$2.73 &   1.59$\pm$  0.21 &  $<$1.67 &  $<$2.47 &   0.36$\pm$  0.09 &   0.52 &  2.7$\pm$0.2 \\
      MRK766 &  0.013 & 13.97$\pm$  0.84 &  20.13$\pm$  1.41 &  15.54$\pm$  1.09 &   9.64$\pm$  0.87 &   0.63$\pm$  0.08 &  13.25 &  2.7$\pm$0.3 \\
      IC3599 &  0.019 &  &   &   &   &   $<$0.07 &   0.05 &  5.2$\pm$1.5 \\
IRAS1322-380 &  0.067 &  6.08$\pm$  0.73 &   7.80$\pm$  0.55 &   3.11$\pm$  0.50 &  $<$2.88 &   1.35$\pm$  0.12 &   1.40 &  4.4$\pm$0.2 \\
      MRK478 &  0.079 &  2.77$\pm$  0.33 &   2.85$\pm$  0.20 &   2.24$\pm$  0.27 &   3.03$\pm$  0.73 &   4.15$\pm$  0.23 &   2.29 &  3.6$\pm$0.1 \\
      MRK493 &  0.032 &  3.88$\pm$  0.47 &   3.47$\pm$  0.21 &   2.30$\pm$  0.28 &   2.20$\pm$  0.46 &   1.56$\pm$  0.11 &   0.78 &  2.7$\pm$0.2 \\
    1652+396 &  0.069 &  &   &   &   &   $<$0.10 &   0.37 &  2.7$\pm$0.3 \\
      AKN564 &  0.024 &  3.41$\pm$  0.82 &   4.13$\pm$  0.37 &   6.78$\pm$  0.47 &  $<$7.47 &   2.77$\pm$  0.17 &   9.17 &  3.4$\pm$0.1 \\
\hline
\end{tabular}

\end{flushleft}
\end{table*}

\section{The ultraviolet spectrum of NLS1}

\subsection{The data}

We have cross--correlated the NLS1 sample discussed by Boller et al. (1996)
against the {\it IUE} database and found that 11 out of 30 NLS1 for which
{\it ROSAT} data are available, had been observed with the short wavelength
spectrograph on {\it IUE}. This subsample was increased by our own
observations of three more NLS1: IRAS~13224-3809, KUG~1031+398 and
Mrk~1044. The log of our {\it IUE} observations is shown in
Table~\ref{log}.  The whole sample is described in Table~\ref{cont}, where
we list the most common object names, the redshifts (from Boller et
al. 1996) and some continuum properties as explained below.

The sample discussed hereafter is in no way statistically
complete; it includes those galaxies that at some stage have been
considered interesting by the scientific community and the {\it IUE} time
allocation committees. 

All the {\it IUE} spectra were re-extracted using the Final Archive
processing software (Nichols et al. 1993). Foreground galactic reddening
has been corrected from the HI column densities map in Dickey \& Lockman
(1990), converted to $E(B-V)$\ according to the relation: $ \langle
N(HI)/E(B-V) \rangle = (6\pm2)\times10^{21}{\rm cm}^{-2}$.  All the spectra
have been redshift corrected to get the wavelength and fluxes in the
objects rest frames.  In order to increase the signal-to-noise ratio, all
the spectra of every single object have been averaged together (see
Sect.~2.2 for a discussion of their variability properties).

The UV continuum has been measured in a 40~\AA\ band centered at 1450~\AA ,
which is apparently free of absorption and/or emission features. Two
(IC3599 and 1652+396) out of the 14 objects were not detected; the
signal-to-noise ratio (S/N) is smaller than 10 in the average spectrum of 5
objects. The average fluxes are given in Table~\ref{cont}.  In this table
we also list the fluxes in the four \iras\ bands from the NED database as
well as the \rosat\ fluxes and spectral slopes reported by Boller et
al. (1996).

\subsection{Variability}

One of the most general properties among normal Seyfert~1 galaxies is
the variability in the UV continuum and broad emission lines. 
The observed UV continuum variations are a few percent on
time scales of a day or less and tens of percent on time scales of
several days. However, 
there is not very much information in the literature about optical and/or
UV variability in NLS1, although they are probably the AGN's which show the
fastest variations in the soft X rays. Only very recently, a sample of 12 NLS1
has been systematically monitored in order to search for optical
variability (Giannuzzo \& Stirpe 1996). Ten of these NLS1 showed
significant variations in the optical continuum and permitted lines over a
time interval of one year.

We have found significant 
changes in the continuum flux in 2 (Mrk~1044 and \irastrece )
out of the 11 detected objects for which more than one spectrum is available. 
Mrk~1044 was observed with \iue\ on December 1 and 20, 1995. Between these
two dates, the continuum increased by 38\%. Unfortunately, in the first SWP
spectrum the peaks of the strongest emission lines (\Lya , CIV) were
saturated. Nevertheless, neither the weaker lines (SiIV, HeII) nor the
wings of \Lya\ and CIV show evidences of variability. 
For Mrk~1044, Giannuzzo \& Stirpe (1996) 
find variations in the \Ha\ and \Hb\ fluxes of --14\%\ and --24\% ,
respectively, between October 1993 and September 1994. They do not
report on the continuum variability. 

There are 11 SWP spectra of 
\irastrece\ from January 1993 to February 1996. 
The variability of the continuum
flux during this time is 24\% (r.m.s.), with a ratio of the maximum to the
minimum flux close to 2.
The analysis of the spectra taken during 1993 showed changes in
the profile of \Lya\ that could be attributed to a variable narrow
absorption (Mas--Hesse et al. 1994). A detailed analysis of the whole data
set for this object is deferred to a later paper.

The number of {\it IUE} observations for the other NLS1 in the sample is
rather small, so that it is not possible to study their variability
properties. 

In summary, the available data suggest that, in the UV, NLS1 can 
vary at least as fast as normal Seyfert~1 galaxies on time scales of
several days. 
With the existing IUE data it has not been possible to 
check whether this class of objects varies faster than normal Seyfert 1s
as in the soft X rays.

\subsection{The continuum spectral energy distribution: 
Comparison with normal Seyfert~1 galaxies}

\begin{table*}
\caption[]{Comparison of NLS1 and normal Seyfert~1 continuum properties}
\label{s1nl}
\begin{flushleft}
\begin{tabular}{lrrrccrrrcr}
\hline
\multicolumn{1}{c}{Parameter} & 
\multicolumn{4}{c}{NLS1 sample$^a$} & &
\multicolumn{4}{c}{Seyfert~1 sample$^b$} &
\multicolumn{1}{c}{Significance$^c$} \\ \cline{2-5} \cline{7-10}
 & 
\multicolumn{1}{c}{N} & \multicolumn{1}{c}{Mean} & \multicolumn{1}{c}{S.D.}
 & \multicolumn{1}{c}{P(Norm.)} & &
\multicolumn{1}{c}{N} & \multicolumn{1}{c}{Mean} & \multicolumn{1}{c}{S.D.}
 & \multicolumn{1}{c}{P(Norm.)} \\
\hline
L$_{100 \mu m}$     & 10 & 44.41 & 0.65 & 0.381 && 21 & 44.36 & 0.79 & 0.032 & 0.787 \\
L$_{ 60 \mu m}$     & 12 & 44.34 & 0.68 & 0.178 && 30 & 44.31 & 0.89 & 0.044 & 0.902 \\
L$_{ 25 \mu m}$     & 10 & 44.37 & 0.56 & 0.896 && 26 & 44.45 & 0.89 & 0.814 & 0.248\\
L$_{ 12 \mu m}$     &  8 & 44.24 & 0.65 & 0.716 && 23 & 44.32 & 0.74 & 0.178 & 0.798 \\
L$_{UV}$            & 12 & 43.74 & 0.77 & 0.839 && 47 & 44.52 & 1.07 & 0.077 & 0.019 \\
L$_{ROSAT}$         & 12 & 43.95 & 0.66 & 0.553 && 53 & 44.42 & 0.93 & 0.188 & 0.106 \\
$\log$(L$_{100 \mu m}$/L$_{ 60 \mu m}$)
                    & 10 & -0.04 & 0.09 & 0.599 && 21 &  0.01 & 0.16 & 0.191 & 0.335 \\
$\log$(L$_{ 25 \mu m}$/L$_{ 60 \mu m}$)
                    & 10 & -0.07 & 0.26 & 0.858 && 26 &  0.08 & 0.30 & 0.531 & 0.202 \\
$\log$(L$_{ 12 \mu m}$/L$_{ 60 \mu m}$)
                    &  8 &  0.00 & 0.26 & 0.547 && 23 &  0.14 & 0.33 & 0.670 & 0.312 \\
$\log$(L$_{UV}$/L$_{ 60 \mu m}$ )
                    & 12 & -0.62 & 0.59 & 0.336 && 27 & -0.07 & 0.73 & 0.260  & 0.028 \\
$\log$(L$_{ROSAT}$/L$_{ 60 \mu m}$)
                    & 12 & -0.39 & 0.64 & 0.441 && 30 & -0.10 & 0.61 & 0.004 & 0.130 \\
$\log$(L$_{UV}$/L$_{ROSAT}$) 
                    & 12 & -0.23 & 0.52 & 0.013 && 47 &  0.11 & 0.35 & 0.219 & 0.028 \\
$\Gamma_{ROSAT}$    & 12 &  3.27 & 0.68 & 0.132 && 53 &  2.44 & 0.47 & 0.690 & $<$0.001 \\
\hline
\end{tabular} \\
$^a$ Boller et al. (1996) sample plus our data,(Sect.~2.1) \\
$^b$ Walter \& Fink (1993) \\
$^c$ Significance of the Mann-Whitney rank sum test on the hypothesis that
  the two samples come from the same parent population (Sect.~2.3)

\end{flushleft}
\end{table*}

\begin{figure}
\epsfxsize=8.8cm
\epsfysize=16cm
\epsffile{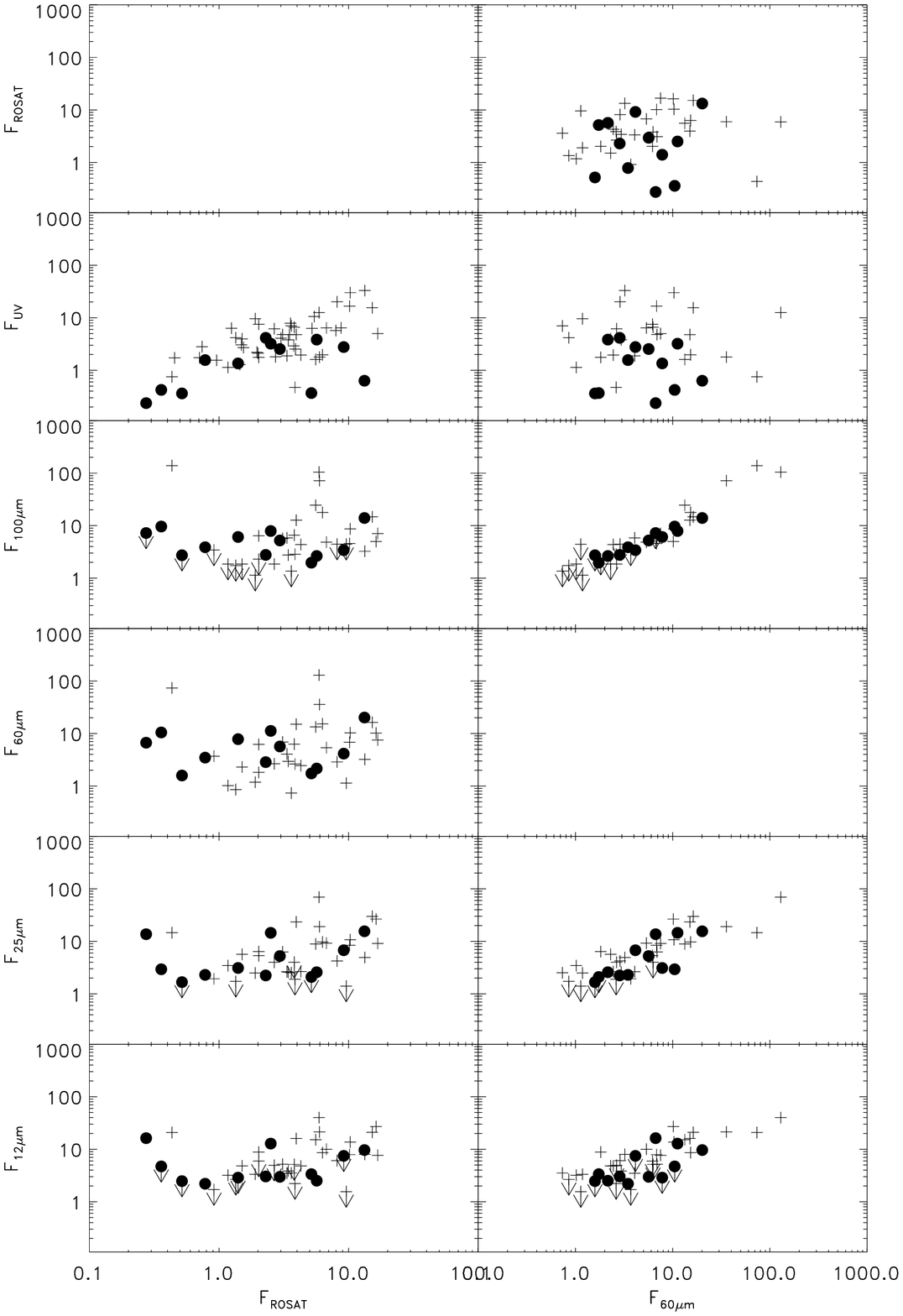}
\caption[]{Flux-flux diagrams for different continuum bands. Note that the
higher energy bands (UV, soft X-rays) are not correlated with lower energy
bands (FIR). Filled circles represent NLS1 galaxies and 
crosses normal Seyfert~1.}

\end{figure}

The spectral energy distribution (SED) of normal broad line Seyfert~1
galaxies is dominated by the ``big blue'' bump (Sanders et al. 1989),
extending from $\sim$4\,000~\AA\ to beyond the shortest observable
wavelengths in the UV region ($\sim$\,1000~\AA). How far this bump extends
into the extreme ultraviolet (EUV) is not known because of the difficulty
of observing extragalactic objects in this spectral region. In this
respect, the ``soft X-ray excess'' found in many AGN 
is sometimes
interpreted as the high energy end of the bump. It is generally thought
that the big blue bump is due to the emission of an accretion disk around a
super-massive black hole, although there are other alternatives proposed
(like free-free emission from optically thin gas clouds or the non-standard
model of emission from a powerful starburst). In all these models, the UV
emission region would be illuminated by the hard X-ray continuum, as required 
by the observed properties of the X-ray and the UV spectra, 
together with their correlated fluctuations. 

Many AGN emit similar amounts of energy in the UV-optical and FIR
regions. There is growing evidence that the nuclear FIR emission in AGN is
due to thermal radiation by dust, although this is still controversial.
Moreover, for some objects it has been argued that the FIR emission is not
directly related to the active nucleus, but to circumnuclear star forming
regions (e.g. Mas--Hesse et al. 1995).  The spectral index between 25 and
60~$\mu$m is a good indicator of the relative contribution of star formation
processes to the FIR emission in AGN (Mas--Hesse et al. 1995); steep FIR
spectra are typical of star forming regions, while flat FIR spectra
indicate the nuclear emission (independently of its nature) dominates.

We have compared the SED of NLS1 with that of normal broad emission line
Seyfert~1 galaxies in order to investigate whether the continuum emission
is originated by the same mechanisms in both types of objects.  The SED
(from the far IR, 100\,$\mu$m, to the soft X-rays, 2.4\,keV) of our sample
of NLS1 galaxies is listed in Table~\ref{cont}.  For comparison we have
selected the sample from the Walter \& Fink (1993) study.  These authors
analyzed the {\it ROSAT All Sky Survey} data of 58 Seyfert~1 galaxies for
which {\it IUE} spectra are available. To compare the X-ray fluxes in both
samples it has been necessary to translate the Boller et al. (1996) 2\,keV
monochromatic fluxes to integrated fluxes in the 0.1--2.4\,keV band, as
given by Walter \& Fink (1993). The translation has been performed using
the soft X-ray spectral indexes given by Boller et al. (1996).

Five out of the 58 Walter \& Fink objects are indeed NLS1, 
also included in the Boller et al. (1996) sample. 
Prior to the comparison of both samples, we have used the reported
properties of the common objects to check the consistency of the two studies. 
We have found differences in the reported soft X-ray spectral
indexes of the five NLS1 of up to 25\% , but within the 3$\sigma$\ error.
Similarly, the fluxes derived by Walter \& Fink (1993) are always larger
than those inferred from the Boller et al. (1996) data.  Nonetheless, in
any case the differences are smaller than 3$\sigma$, according to the
errors given by Walter \& Fink (1993). Therefore, we can be confindent that
the results of both studies are consistent within their errors bars.  The 5
NLS1 in Walter \& Fink (1993) have been excluded of their sample, leaving a
set of 53 normal broad line Seyfert~1 galaxies. We also recall here that
the Boller et al. (1996) sample of NLS1 galaxies has been enlarged with
three more objects observed by us with {\it IUE} (Sect.~2.1).

In Table~\ref{s1nl} we give the mean and standard deviation (S.D.) of some
parameters in both samples, as well as the number of objects for which each
parameter is available and the probability that the parameter is normally
distributed. The luminosities in the \iras\ bands, in the UV (1450~\AA) and
in the \rosat\ band, as well as the ratios between some of them are
given. The last column in the table is the significance of the hypothesis
that both samples come from the same parent population according to the
Mann-Whitney rank sum test.

The smallest significance (i.e., the highest probability that the two
samples come from different parent distributions) is found for the spectral
index in the \rosat\ band ($\Gamma$). However, it should be noted that the
S.D. of $\Gamma$\ is larger in the sample from Boller et al. (1996); many
NLS1 have $\Gamma$\ well within the typical values found for normal broad
emission lines Seyfert~1, but there are some NLS1 that show much steeper
soft X-ray spectra. Therefore, a very steep soft X-ray spectrum is not a
characteristic of all NLS1, although the steepest spectra among Seyfert~1
galaxies are found in NLS1. In spite of this difference in spectral slopes,
there is no statistically significant (significance $\le$ 0.10) difference in
the total \rosat\ luminosity between NLS1 and normal Seyfert~1 galaxies. 

The other parameters that are different in the two samples (at the 0.05
significance level) are the UV luminosity and the ratios in which this
luminosity is involved. The NLS1 are, on average, 6 times fainter in the UV
than normal broad emission lines Seyfert~1, although the total \Luv\ range
spanned by the NLS1 ($10^{42-45}$\ergs ) is well within the total \Luv\
range for normal Seyfert~1 galaxies ($10^{41.5-46.5}$\ergs ). The average
\Luv/\Lx\ and \Luv/\L60\ ratios are correspondingly smaller in NLS1 than in
normal Seyfert~1; furthermore, the smallest absolute values of these ratios
correspond to NLS1. It is also worth noting that the ``normal'' Seyfert~1
with \Luv/\Lx\ and \Luv/\L60\ closer to the extreme values found for NLS1
are NGC~4051 and NGC~1566. Winkler (1992) gives a FWHM for the 
\Hb\ line in NGC~1566 of 1800 km/s and 
Filippenko \& Sargent (1985) describe the Balmer
lines of NGC~4051 as 
``not very broad compared with those in most Seyfert~1
galaxies''. 

From Table~\ref{s1nl}, it is remarkable the high degree of similarity in
the FIR properties between NLS1 and normal broad emission lines Seyfert~1.
This result is consistent with the Halpern \& Oke (1987) results.  The
spectral index between 25 and 60~$\mu$m (\afir ) of the five NLS1
discussed by Halpern \& Oke (1987) spans the whole range found for normal
Seyfert~1 galaxies (Miley et al. 1985, Mas--Hesse et al. 1995). For our
sample, we find that only two out of ten NLS1 detected at 25 and 60~$\mu$m
have \afir\ steeper than --1.5, typical of star forming regions. Although
our sample is rather small, it is still worthy to note that the fraction of
steep FIR spectrum NLS1 is roughly consistent with that found among normal
Seyfert~1 by Mas--Hesse et al. (1995).

We have compared the luminosity in all the \iras , UV and \rosat\
bands {\it versus} \Lx\ and \L60. We find a statistically significant
correlation among the luminosities in all {\it IRAS}, UV and {\it ROSAT}
bands. Moreover, there is no statistical evidence for the NLS1 and
Seyfert~1 galaxies to show different slopes in the linear regression
fits. However, the positive correlations found in all continuum bands are
not held when the fluxes, instead of luminosties, are considered
(Fig.~1). When the distance effect is removed, the correlations between low
energy (\iras) and high energy (UV, \rosat) bands disappear, leaving only
the correlation among \iras\ bands on one side and between the UV and
\rosat\ bands on the other side.  These results suggest that the mechanisms
producing the UV and soft X-ray photons are strongly related, but the
connection between the production of FIR radiation and UV -- soft X-rays is
not straightforward.

The conclusion that emerges from the comparison of the luminosity in the
FIR, UV and soft X rays continuum bands is that the SED of NLS1 and normal
broad emission lines Seyfert~1 galaxies are very similar, except in that
some, but not all, NLS1 have steeper soft X-ray spectra and that NLS1 tend
to be somewhat underluminous in the UV region.

\subsection{Line Profiles}

A first look to the line profiles suggests the presence of broad
wings in the strongest emission lines (Fig.~2). In order to test it, we
have fitted the \Lya , \CIV\ and \HeII\ line profiles with first only one
and then two gaussian components (for the \Lya\ profile an additional
gaussian has been included to account for \NV ). The best fit parameters
for the case that results in a smaller reduced chi-square\footnote{We have
used the minimization package {\it MINUIT} from the CERN Library, James
(1994)} (\chicuar = \chicua/$\nu$, where $\nu$\ is number of degrees of
freedom) are shown in Table~\ref{line}. For the seven objects where the
signal-to-noise ratio (S/N) in the continuum is larger than 10, we note
that the multi-component fit is preferred in at least one of the emission
lines. Moreover, when the line profile is better fitted by two gaussians,
we find that one is slightly broader than the \iue\ Point Spread Function
(FWHM$\sim$1500\,km/s) and the other has a FWHM similar to those found in
normal Seyfert~1 galaxies (FWHM~$\ge$~5000\,km/s). For the rest of the
objects, the multi-component fit does not improve with respect to the fit
with a single gaussian. However, we point out that the spectra of these
objects are those which have the lowest S/N making it difficult to detect
broad line wings. In any case, the line widths for the single-component
fits tend to be larger than the width of the narrow component in the
multigaussian fits.  We want to stress that it is not our aim to identify
each gaussian component in the fits with physically different regions. The
purpose of the fits is to confirm or reject the presence of broad wings in
the emission line profiles.

Our results strongly suggest the presence of high velocity line emitting
gas in the nuclei of NLS1 and that the profiles are qualitatively similar
to those found in normal Seyfert~1, in the sense that they can be roughly
characterized by a narrow plus a broad component.

As mentioned above, NLS1 as a class are characterized by optical hydrogen 
emission lines broader than their neighbour forbidden lines, but narrower
than the Balmer lines found in normal Seyfert~1 galaxies (Osterbrock \&
Pogge, 1985). Nevertheless, after finding evidence of broad wings
(FW0I$\ge$10,000 km/s and FWHM$\sog$5,000 km/s ) in the permitted UV 
lines of some NLS1, we have
specifically searched for such broad wings in the hydrogen  optical
lines. In their variability analysis of NLS1 galaxies, Giannuzzo \& Stirpe
(1996) present \Ha\ and \Hb\ spectra of three NLS1 objects for which we find
broad wings in the UV lines (Mrk~359, Mrk~1044 and Akn~564). 
To check the presence of broad wings, we have done a multi-gaussian fit 
(similar to that performed with the UV lines) to these spectra, kindly
provided by Giannuzzo \& Stirpe.
The fits require, first, three narrow gaussian lines whose width is
determined by the spectral resolution: one gaussian for the Balmer line
(\Hb\ or \Ha ) in the range and two additional ones to account for either
the \Oiii\ or the \Nii\ doublet. In the \Hb\ range, two more gaussians,
slightly broader (i.e. $\sim$ 1000 km/s), are included in the fits to
account for the FeII emission in Akn~564 and Mrk~1044.
In addition to that, both \Ha\ and \Hb\ in the three objects require at
least two more, broader, gaussian components to account for the total
profile.  The first of them is of intermediate width, $\sim 1000$ km/s,
and the second has an FWHM less than 3000 km/s,
narrower than the broadest UV component in the same objects.  In order to
check the existence of broader wings, we have tried another fit in which
the width of the broadest line is fixed to 5000 km/s. The result is clearly
worse as the r.m.s of the residuals is much larger. Therefore, we can only
put upper limits to the flux of a putative optical component of 5000~km/s
FWHM, as it is certainly not detected in the available spectra. These
estimates can then be used to infer the upper limit of the broad \Lya/\Ha\
ratio.  From the residuals of the multigaussian fits we obtain that, for
the three NLS1 studied, the broad \Lya/\Ha\ ratio is larger than 100.
This lower limit is also confirmed when directly estimated from the S/N in
the wings of the lines.

\begin{table*}
\caption[]{Results of line profile fits to the UV emission lines}
\label{line}
\begin{flushleft}
\begin{tabular}{lcccccc}
\hline \multicolumn{1}{c}{Object} & \multicolumn{2}{c}{Ly$\alpha$} &
\multicolumn{2}{c}{CIV} & \multicolumn{2}{c}{HeII} \\ & Narrow & Broad &
Narrow & Broad & Narrow & Broad\\ & Flux$^{a}$ & Flux$^{a}$ & Flux$^{a}$ &
Flux$^{a}$ & Flux$^{a}$ & Flux$^{a}$ \\ & Width$^{b}$ & Width$^{b}$ &
Width$^{b}$ & Width$^{b}$ & Width$^{b}$ & Width$^{b}$ \\ \hline \\ MRK957 &
& & & & & \\ & & & & & & \\ \\ IZW1 & 114$\pm$28 & 239$\pm$82 & 42$\pm$16 &
53$\pm$52 & 4.5$\pm$4.3 & 9.5$\pm$6.2 \\ & 1760$\pm$240 & 6800$\pm$1200 &
2200$\pm$430 & 6600$\pm$3400 & 1020$\pm$510 & 9100$\pm$3500 \\ \\ MRK359 &
75$\pm$28 & 100$\pm$150 & 53$\pm$13 & 45$\pm$25 & 12.0$\pm$4.2 & \\ &
2220$\pm$410 & 8900$\pm$8400 & 1830$\pm$210 & 4820$\pm$860 & 1840$\pm$330 &
\\ \\ MRK1044 & 308$\pm$41 & 143$\pm$49 & 69$\pm$15 & 104$\pm$35 &
9.6$\pm$3.9 & 52$\pm$16 \\ & 2410$\pm$150 & 10900$\pm$2400 & 2150$\pm$250 &
6130$\pm$750 & 2000$\pm$390 & 10800$\pm$2000 \\ \\ MRK1239 &
\multicolumn{2}{c}{41.5$\pm$14.8} & \multicolumn{2}{c}{22$\pm$11} &
\multicolumn{2}{c}{11.5$\pm$8.1} \\ & \multicolumn{2}{c}{2960$\pm$570} &
\multicolumn{2}{c}{3310$\pm$880} & \multicolumn{2}{c}{1960$\pm$800} \\ \\
KUG1031+398 & 12.2$\pm$4.2 & 15$\pm$12 & & & & \\ & 1590$\pm$310 &
7400$\pm$3000 & & & & \\ \\ MRK42 & \multicolumn{2}{c}{23.8$\pm$6.0} &
19.6$\pm$6.1 & 6.0$\pm$7.3 & 3.8$\pm$2.8 & \\ &
\multicolumn{2}{c}{3050$\pm$410} & 2320$\pm$430 & 3100$\pm$2400 &
1360$\pm$520 & \\ \\ MRK766 & \multicolumn{2}{c}{16.4$\pm$3.9} &
\multicolumn{2}{c}{15.3$\pm$4.6} & \multicolumn{2}{c}{9.8$\pm$4.5} \\ &
\multicolumn{2}{c}{1740$\pm$220} & \multicolumn{2}{c}{2910$\pm$440 } &
\multicolumn{2}{c}{3400$\pm$870} \\ \\ IC3599 & & & & & & \\ & & & & & & \\
\\ IRAS1322-3809 & 13$\pm$26 & 12$\pm$26 & \multicolumn{2}{c}{16.7$\pm$9.1}
& \multicolumn{2}{c}{9.5$\pm$7.3} \\ & 1900$\pm$1400 & 3000$\pm$3800 &
\multicolumn{2}{c}{6500$\pm$2000} & \multicolumn{2}{c}{7200$\pm$3400} \\ \\
MRK478 & 149$\pm$29 & 125$\pm$62 & 30$\pm$15 & 64$\pm$34 &
\multicolumn{2}{c}{46$\pm$17} \\ & 2030$\pm$210 & 5900$\pm$1200 &
1680$\pm$460 & 5300$\pm$1100 & \multicolumn{2}{c}{8900$\pm$2000} \\ \\
MRK493 & 44$\pm$28 & 53$\pm$53 & 24$\pm$13 & 30$\pm$27 & 4.9$\pm$4.3 & \\ &
1850$\pm$460 & 4000$\pm$1300 & 2110$\pm$510 & 5300$\pm$1500 & 2200$\pm$1100
& \\ \\ 1652+396 & & & & & & \\ & & & & & & \\ \\ AKN564 &
\multicolumn{2}{c}{120$\pm$15} & \multicolumn{2}{c}{28$\pm$11} &
16.9$\pm$5.3 & 31$\pm$17 \\ & \multicolumn{2}{c}{2490$\pm$160} &
\multicolumn{2}{c}{3840$\pm$770} & 1460$\pm$240 & 9850$\pm$2750 \\ \hline
\end{tabular}

Notes: \\ $^{a}$: $10^{-14}$\ergcms \\ $^{b}$: FWHM (km/s)
\end{flushleft}
\end{table*}

\begin{figure}
\epsfxsize=8.8cm \epsfysize=16cm
\epsffile{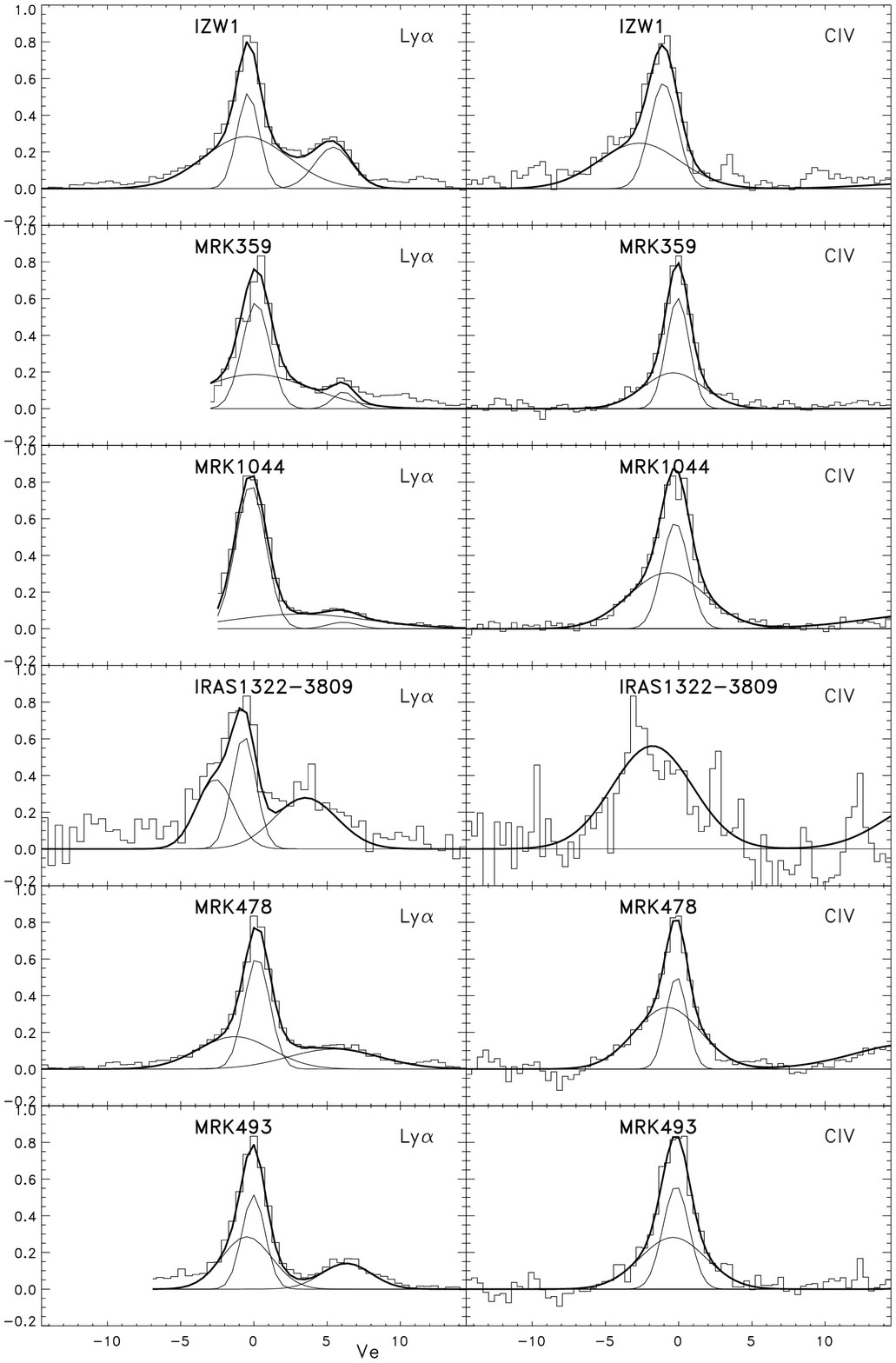}
\caption[]{\Lya\ and CIV profiles of some NLS1 observed with \iue , showing
broad wings similar to those found in normal broad line Seyfert~1
galaxies. Thin lines show the gaussian components discussed in the text,
while the thick line shows the total fit.}
\end{figure}

\section{Discussion}

Independently of the details on how the photoionizing radiation is
produced, the intensity of the emission lines and their response to
variations in the continuum strongly argues in favour of the hypothesis
that there is photoionized gas in the nuclei of active galaxies. In the
standard model of AGN, such a gas is distributed in a large number of
relatively small clouds surrounding an accretion disk around a supermassive
black hole, whose emitted radiation heats and ionizes the gas clouds. The
intensities and profiles of the emission lines are driven by the
photoionization continuum and the physical conditions, distribution and
kinematics of the gas. As mentioned in Sect.~1, the lack of broad wings in
the optical permitted lines in NLS1 has been attributed to either the
absence of high velocity gas clouds or an orientation effect that makes
that these clouds have a negligible velocity component along the line of
sight. However, the detection of broad wings in the permitted UV lines
proves the presence of high velocity gas clouds, with a large velocity
component along the line of sight.  Therefore, it is necessary to conceive
a scenario where a central continuum similar to that of the normal
Seyfert~1 ionizes high velocity gas which emits broad UV lines that can be
seen from the Earth, whereas the broad optical emission lines can not be
detected.

A possibility is that both UV and broad optical lines are in fact emitted
in the BLR, but hidden by the thick molecular dusty torus proposed in the
unification models for AGN. The broad UV lines then will show up after
reflection in a scatterer material above the torus. In this context, a
natural way to show up the UV but not the optical lines is scattering by
dust, which is much more efficient in the UV than in the optical (as
opposed to scattering by electrons, which is essentially wavelength
independent).  
If this picture is true and the BLR is hidden from direct view, we would
expect the continuum to be hidden as well, unless the system has a very
unlikely geometry. However, the SED of NLS1 discussed in previous section
does not support the picture of hidden AGN.  In fact, the NLS1 SED is
closer to that of normal Seyfert~1 galaxies than to that of Seyfert~2 (up
to now the best candidates to hidden AGN). In particular, the Seyfert~2 SED
is much more strongly peaked in the FIR range than that found in NLS1.
Moreover, another observational result in conflict with the hypothesis of a
hidden AGN is the absence of significant cold absorption of the X-rays
(Boller et al. 1996).

There is another possibility that does not consider a hidden nucleus, but a
scenario in which the nuclear gas simply does not emit optical lines.  The
strongest permitted lines in the optical range are \Ha , \Hb\ and the rest
of the lines from the Balmer series of hydrogen. These lines are
predominantly formed in regions where the hydrogen is only partially
ionized (HPI), while the strongest UV lines (\CIV\ and \HeII) form where
the hydrogen is fully ionized (HFI) (H-\Lya\ may form in both fully or
partially ionized regions). Therefore, the absence of broad H-Balmer lines
in the optical region would be naturally explained if there is no HPI zone
in the high velocity gas in the nuclei of NLS1.  The presence within the
BLR of hydrogen fully ionized clouds (and, hence, optically thin to Lyman
continuum photons), has already been suggested by Shields et al. (1995)
(see also references therein).  As they show, thin clouds can contribute
significantly to the emission of the broad high ionization lines in AGN
without producing significant low ionization species. Shields et al. (1995)
also discuss on the possibility that these thin clouds are responsible of
the observed UV and soft X-ray absorption features in the AGN spectra (when
crossing our line of sight to the central source).

In order to explore the physical conditions of a gas to emit the observed
NLS1 line spectrum, we have made use of the photoionization code {\it
CLOUDY} (Ferland 1991). We have assumed a solar abundance gas which is
illuminated by an ionizing continuum similar to that derived by Mathews \&
Ferland (1987).  The free parameters (the gas physical conditions) are
selected automatically by the code to find an optimal solution fitting the
observed average line spectrum: \Lya/CIV$\sim$2, \Lya/\Hb~$\ge$~100,
\HeII/CIV$\sim$0.10. The code derives an optimal model for a region where
the total hydrogen column density is $<10^{20}\,{\rm cm}^{-2}$, the
hydrogen number density is $\le\,10^{7.5}\,{\rm cm}^{-3}$\ and the
ionization parameter $\log (U)\,\ge\,-0.5$. Under these conditions, the
emitting gas is fully ionized in hydrogen, and, thus, optically thin to the
Lyman continuum.  We note that the detailed physical conditions would be
different if, for instance, the ionizing continuum is harder than that of
Mathews \& Ferland (1987). However, we want to stress that, independtly of
the actual numbers obtained from {\it CLOUDY}, a high ionization parameter
and a relatively low column density are required to roughly reproduce the
observed line ratios.  A detailed model of the BLR in NLS1 is out of the
scope of this paper, since higher quality data over a wider spectral range
would be needed for every single object, but the {\it CLOUDY} computations
give sufficiently general results for this dicussion.

The equivalent width of the \Lya\ line emitted in optically thin clouds is
much smaller ($\ga$100 times) than in the optically thick case, but the
observed equivalent width of the broad \Lya\ in NLS1 (as derived from
Tables~\ref{cont} and \ref{line}) is only $\sim$3 times smaller than in
normal Seyfert~1 galaxies ($\sim 100$\,\AA ).  If the BLR in NLS1 emits
indeed in an optically thin regime, the covering factor
($\Omega\,\equiv$\,fraction of the sky covered by clouds, as seen by the
central continuum source) has to be much larger ($\Omega\,\la\,1$) than in
normal Seyfert~1 ($\Omega\,\sim\,0.01$) to keep the difference in
equivalent widths not too big.  As a side effect, a large value of
$\Omega$\ implies that the probability of finding a cloud right in the
observer's line of sight is very high. However, the effect of an optically
thin intervening cloud in the observed spectrum will not be very strong,
except in the soft X-ray domain, which is very sensitive to relatively
small amounts of neutral hydrogen and where absorption edges can be found
due to some highly ionized atoms (e.g., OVI-VIII). As we noted above, this
fact was already pointed by Shields et al. (1995) who discuss under which
conditions the thin BLR clouds can also produce soft X-ray absorption
edges.  In this respect, we note that the crossing time of one of these
optically thin clouds would be of the order of hours, similar to the
variability time scales found in the soft X rays of most NLS1.  If the
variations detected in the soft X-rays are due to the passage of individual
clouds across the observer's line of sight, no change in the emission lines
would be expected, since their flux comes from the integration over a very
large number of clouds, whose ionizing radiation has not changed.

Another implication of optically thin clouds is that the response of the
emission lines to continuum variations is much weaker than in the case of
thick clouds. A line fluctuation would only be detected if the change in
the continuum is sufficiently large to produce a significant change in the
ionization structure of the clouds. This could explain why the wings of the
UV lines in the spectra of Mrk~1044 and \irastrece\ remained constant while
the continuum and the line cores varied (Sect.~2.2).

In their study of optically thin broad-line clouds in AGN, Shields et
al. (1995) suggest that the importance of thin clouds relative to the thick
clouds should be larger in low luminosity objects than in high luminosity
ones.  To account for this effect, they claim for outflows of the thin
clouds that would proceed more efficiently in intrinsically brighter
sources.  This could explain the observed anti-correlation between UV
emission-line equivalent width and continuum luminosity (the Baldwin
effect).  In this context, the tendency for the NLS1 galaxies to have lower
UV luminosities than ``normal'' Seyfert~1 also supports the presence of
optically thin broad-line clouds in NLS1.

We have finally compared the ``typical'' parameters that characterize
normal BLR ($\log(U)=-2\; ; n_{H}\sim 10^{10}{\rm cm}^{-3}$), with those
found to represent the BLR in NLS1 (($\log(U)=-0.5 ; n_{H}\sim 10^{7.5}{\rm
cm}^{-3}$). From the definition of $U$,
\begin{equation}
U\,=\frac{\int_{\nu_{0}}^{\infty}\frac{L(\nu)}{h\nu}d\nu} {4\pi r^{2} c
n_{H}}
\end{equation}
where $\nu_{0}$ corresponds to 13.6~keV, and taking into account the
average differences in \Luv\ and \Lx\ given in Table~\ref{s1nl}, the
distance $r$ of the BLR to the ionizing source should be very similar in
NLS1 (within a factor of 2) and normal Seyfert~1 galaxies.

\section{Summary}

We have analyzed the UV properties of a sample of the 14 NLS1 galaxies
observed with the {\it IUE} satellite. This sample comprises some
unpublished observations of our own together with archival data. When the
available number of spectra and their S/N allowed for it, we have studied
the UV variability properties of the objects. Two objects (Mrk~1044 and
\irastrece ) have been found to experience significant variations in the
continuum while the emission line wings remain constant.

We have studied the continuum SED of the NLS1 sample, from the FIR to the
soft X-rays, comparing it with the SED of a sample of ``normal'' Seyfert~1
galaxies from Walter \& Fink (1993).  The FIR properties of NLS1 are not
different from the FIR properties of ``normal'' Seyfert~1 galaxies with
broad optical permitted lines. Only the UV luminosities tend to be smaller
in NLS1, and the soft X rays spectra tend to be steeper, although a number
of NLS1 are found to have spectral indexes very similar to those found in
``normal'' Seyfert~1 galaxies.

The analysis of the UV emission line spectra of the sample, has revealed
that all NLS1 galaxies detected with \iue\ at sufficiently high S/N show
broad wings in their emission lines (\Lya , CIV and HeII). The positive
detection of these broad wings demonstrates the existence of gas with a
large velocity component in our line of sight, ruling out the hypotheses
that (a) there is no BLR in NLS1, or (b) there is a BLR, but we cannot see
it due to orientation effects. It is also difficult to accept that the
broad wings we see in the UV emission lines result from a scattering
process, since the optical counterparts are not detected even in polarized
light and since the NLS1 continuum SED does not suggest a hidden AGN as in
Seyfert~2 galaxies.

We have discussed how the absence of broad optical lines can be explained
if the partially ionized zone in the high velocity line emitting gas is
missing. The BLR would be made of fully ionized hydrogen gas in optically
thin clouds similar to those proposed by Shields et al. (1995).  In this
way only the high ionization lines would be produced.  Some of the
implications of a BLR formed by highly ionized gas are discussed.  In
particular, the covering factor needed to account for the observed
equivalent width of the emission lines should be very large.

\acknowledgements{We are grateful to G. Stirpe and E. Gianuzzo for
providing their optical spectra of three NLS1. We also want to thank
T. Boller for his help in the selection of the targets list. This work has
been partially supported by Spanish CICYT grant PB-ESP95-0389-C02-02.}

\end{document}